\documentstyle[twocolumn,aps]{revtex}

\begin{document}
\title{Scaling and Crossover to Tricriticality in Polymer Solutions}
\author{M. A. Anisimov$^{1}$, V. A. Agayan$^{1}$, and E. E. Gorodetskii$^{1,2}$}
\address{$^{1}$ Institute for Physical Science and Technology and Department of\\
Chemical Engineering, \\
University of Maryland, College Park, MD 20742, U.S.A.\\
$^{2}$ Oil and Gas Research Institute of the Russian Academy of Sciences,\\
Gubkina 3, 117971 Moscow, Russia}
\date{\today}
\maketitle

\begin{abstract}
We propose a scaling description of phase separation of polymer solutions.
The scaling incorporates three universal limiting regimes: the Ising limit
asymptotically close to the critical point of phase separation, the ``ideal
gas'' limit for the pure-solvent phase, and the tricritical limit for the
polymer-rich phase asymptotically close to the {\it theta} point. We have
also developed a phenomenological crossover theory based on the
near-tricritical-point Landau expansion renormalized by fluctuations. This
theory validates the proposed scaled representation of experimental data and
crossover to tricriticality.

\end{abstract}


Phase separation in solutions of polymers in low-molecular-weight
(monomer-like) solvents changes dramatically with increase of the degree of
polymerization (Fig. 1) \cite{DK:80}. Like in simple binary liquids,
asymptotically close to the critical point the coexistence curves obey a
universal power law of the form
\begin{equation}
\phi -\phi _{{\rm c}}=\pm B_{0}\left| \tau \right| ^{\beta },
\label{phi_asympt}
\end{equation}
where $\phi $ is the volume fraction of polymer, $\phi _{{\rm c}}$ is the
critical volume fraction; $\tau =(T-T_{{\rm c}})/T$, $T$ is the temperature,
$T_{{\rm c}}$ is the critical temperature; $\beta =0.326\pm 0.001$ is a
universal 3-dimensional (3d) Ising critical exponent \cite{GZ:87}, $B_{0}$
is a system-dependent critical amplitude. However, with increase of the
polymer molecular weight, the range of validity of the symmetric
parabolic-like behavior given by Eq.(\ref{phi_asympt}) shrinks, yielding an
asymmetric angle-like coexistence boundary near the {\it theta} point \cite
{Fl:53} (Fig. 1). Physically, it means that in the limit of infinite
molecular weight (upon approaching the {\it theta} point) the critical
amplitude $B_{0}$ and the range of 3d-Ising behavior vanish.

Qualitatively, the phenomenon of separation of a polymer solution into two
coexisting phases was explained long ago by Flory \cite{Fl:53}. According to
the Flory theory, the dependence of the critical temperature $T_{{\rm c}}$
and the critical volume fraction $\phi _{{\rm c}}$ of the polymer on the
degree of polymerization $N$ is $T_{{\rm c}}=\Theta /(1+1/\sqrt{N})^{2}$ and
$\phi _{{\rm c}}=1/(1+\sqrt{N})$, where $\Theta $ is the {\it theta}
temperature. As elucidated by Widom \cite{Wi:93}, for any value of the
scaling variable $x=\frac{1}{2}\tau \sqrt{N}$ (where $N$ is assumed to be
large and $\tau $ to be small), the phase coexistence in the Flory system
can be represented in terms of a scaling form. The concentration difference $%
\phi ^{\prime \prime }-\phi ^{\prime }$, where $\phi ^{\prime \prime }$ and $%
\phi ^{\prime }$ are the volume fractions of polymer in the concentrated and
dilute phases, respectively, is given by
\begin{equation}
\sqrt{N}(\phi ^{\prime \prime }-\phi ^{\prime })\sim \left\{
\begin{array}{ll}
2\sqrt{6x} & (x\rightarrow 0) \\
3x & (x\rightarrow \infty )
\end{array}
\right. .  \label{Widom_scaling}
\end{equation}
Although Eq. (\ref{Widom_scaling}) yields the angle-like coexistence in the
{\it theta-}point limit ($x\rightarrow \infty $), it violates Eq. (\ref
{phi_asympt}) in the critical-point limit ($x\rightarrow 0$). The reason is
well known: the Flory theory is essentially a mean-field theory which, just
like the van der Waals theory for simple fluids, ignores critical
fluctuations. It is possible to modify the Flory model to include critical
fluctuations, which does indeed predict both the critical and the {\it theta}%
-point limits correctly as well as the crossover between them \cite{PS:99}.
However, restrictions implied by the Flory model for the system-dependent
parameters (even after incorporating the fluctuations), and, especially, for
the dependence of these parameters upon the degree of polymerization, are
too tight to apply the model to real systems.

An attempt to describe the data shown in Fig. 1 by a generalized form of Eq.
(\ref{Widom_scaling}) with $\sqrt{N}$ replaced by $\phi _{{\rm c}}^{-1}$,
the $\sqrt{x}$ by $x^{\beta }$ at the limit $x\rightarrow 0$, and $x$ by $%
(T_{{\rm c}}-T)/(\Theta -T_{{\rm c}})$ in both limits, was made by Isumi and
Miyake \cite{IM:84}. A practical disadvantage of this approach is that, when
$T_{{\rm c}}$ is close to the {\it theta} temperature (large $x$), even
small changes in $\Theta $ (which is not a directly obtainable parameter)
cause dramatic changes in $x$, making the scaling representation extremely
sensitive to the choice of $\Theta $.

In this letter we propose a general scaling description of phase separation
in polymer solutions. Experiments have shown that $\phi _{{\rm c}}$ does not
satisfy the dependence on the degree of polymerization implied by the Flory
theory \cite{Sa:89}. The description we propose is not based on any specific
molecular model and does not incorporate any particular dependence of the
critical parameters on the degree of polymerization. Instead, it uses
experimentally well defined variables, namely, the reduced temperature
distance to the critical point $\tau $ and the critical volume fraction $%
\phi _{{\rm c}}$. Furthermore, to obtain an explicit form of the scaling
function, we have developed a crossover theory by incorporating fluctuations
into a Landau expansion near the tricritical point. Furthermore, we shall
elucidate the physical nature of the crossover phenomena: very close to the
critical phase-separation point the correlation length of the concentration
fluctuations becomes much larger than the polymer molecular size (radius of
gyration) and the system exhibits universal 3d-Ising behavior. Very close to
the {\it theta} point the radius of gyration becomes larger than the
correlation length and the system exhibits tricritical mean-field behavior
\cite{MW:97}.

We assume that a polymer solution can be described by a scaling function $%
y(z)$ with three universal limits
\begin{equation}
y(z)=\left\{
\begin{array}{cc}
\pm Kz^{\beta } & (z\rightarrow 0), \\
\frac{1}{2}z & (z\rightarrow \infty ,\text{ polymer-rich phase}), \\
1 & (z\rightarrow \infty ,\text{ solvent-rich phase}),
\end{array}
\right.  \label{3limits}
\end{equation}
where
\begin{equation}
y=A(\phi -\phi _{{\rm c}})/B_{0}\phi _{{\rm c}}^{\beta },\text{ \ \ \ \ \ }%
z=C\left| \tau \right| /\phi _{{\rm c}},  \label{ybest}
\end{equation}
and $K=AC^{-\beta }$ with $A$ and $C$ being system-dependent coefficients.
The coefficient $C$ defines the limiting $(M_{{\rm w}}\rightarrow \infty )$
slope of the phase-separation boundary (Fig. 1). The coefficient $A$ can be
obtained from a linear correlation between the asymptotic amplitude $B_{0}$
and $\phi _{{\rm c}}^{1-\beta }$ (insert in Fig. 1) for high molecular
weights of polymer (small $\phi _{{\rm c}}$), so that $y=(\phi -\phi _{{\rm c%
}})/\phi _{{\rm c}}$ in this limit. The coefficient $A$ becomes a weak
function of $\phi _{{\rm c}}$ for lower molecular weights and thus allows
for incorporating non-asymptotic regular effects. The Ising limit in Eq. (%
\ref{3limits}) will be perfectly universal for different systems if the
coefficient $K=A/C^{\beta }$ is not system-dependent. Although there is no
theoretical proof for such universality, for the three polymer solutions we
have analyzed, the combination $AC^{-\beta }$ turns out to be the same.

In Fig. 2 we show coexistence-curve data obtained by Dobashi {\it et al. }
\cite{DK:80} for polystyrene in methylcyclohexane, by Xia {\it et al.} \cite
{XS:96} for polymethylmethacrylate in 3-octanone, and by Nakata {\it et al.}
\cite{NC:78} for polystyrene in cyclohexane, scaled according to Eq. (\ref
{3limits}). We see that all data points collapse onto a single master curve.
In Fig. 4, a crossover from critical Ising behavior (for $z\ll 1$) to the
{\it theta} behavior (for $z\gg 1$) is clearly seen. As $z$ increases, the
volume fraction $\phi ^{\prime }$ of the solvent-rich phase goes to zero
(the ``ideal-gas'' limit), while the volume fraction $\phi ^{\prime \prime
}/\phi _{{\rm c}}$ of the polymer-rich phase tends to its theta limit $\frac{%
1}{2}z$, indicated in Fig. 4 by the dashed line. The slope of the dashed
line on a double logarithmic scale corresponds to the tricritical value of $%
\beta =1$.

De Gennes \cite{deGennes} has pointed out that the {\it theta} point in the
polymer-solvent system is a tricritical point. A tricritical point is a
point which separates lines of second-order ($\lambda $-line) and
first-order transitions. The states above the {\it theta} temperature on the
$\phi =0$ (shown by the cogged line in Fig. 1) correspond to the
critical-like self-avoiding-walk singularities associated with the behavior
of long $(N\rightarrow \infty )$ polymer molecules at infinite dilution \cite
{deGennes,F:94}. This $\lambda $-line is associated with an $n$-component
vector order parameter ($\psi $) in the limit $n\rightarrow 0$ \cite
{deGennes}. The field $h$, conjugate to the order parameter, is zero along
the $\lambda $-line but it becomes non-zero for finite degrees of
polymerization. The correlation length associated with the order parameter
is the radius of gyration, which diverges in the limit of infinite degree of
polymerization (zero field). Below the {\it theta} (tricritical) point, the
polymer order parameter exhibits a discontinuity accompanied by phase
separation and by a discontinuity in the concentration of the polymer. The
line of critical phase-separation points shown in Fig. 1 is a nonzero-field
critical (``wing'') line originating from the tricritical point. The order
parameter for the fluid-fluid phase separation, associated with the
concentration $\phi $, and the polymer order parameter $\psi $ belong to
different classes of universality. Tricriticality emerges as a result of a
coupling between these two order parameters and exhibits mean-field behavior
with small logarithmic corrections \cite{S:75}.\ Physically, $\psi $ is
proportional to the concentration of end points of the polymer chain, while
the concentration $\phi $ is proportional to $\left| \psi \right| ^{2}$ \cite
{deGennes}. Therefore, a proper description of the phase separation near the
tricritical point should incorporate a crossover between Ising critical
behavior and (almost) mean-field tricritical behavior.

To obtain an explicit form of the proposed scaling description, we start
with the Landau expansion of the critical part of the dimensionless
thermodynamic potential $\Delta \tilde{\Omega}$ of a two-component system in
the vicinity of the tricritical point in powers of the order parameter $\psi
$ \cite{LL:80}:

\begin{equation}
\Delta \tilde{\Omega}=\tilde{\tau}\psi ^{2}-\lambda \psi ^{4}+v\psi
^{6}-h\psi ,  \label{exp_phi}
\end{equation}
where $h$ is the ordering field; $\tilde{\tau}=a(\Delta \tilde{\mu}+b\Delta
\tilde{T})$ is the temperature-like scaling field, $\Delta \tilde{T}%
=(T-\Theta )/T$ with $\Theta $ being the tricritical ({\it theta})
temperature, $\Delta \tilde{\mu}=\tilde{\mu}-\tilde{\mu}_{\Theta }$ with $%
\tilde{\mu}=(\mu _{2}/v_{2}-\mu _{1}/v_{1})/RT$ being the reduced
polymer/solvent chemical potential difference, $v_{2}$ and $v_{1}$ the
corresponding molecular volumes, $\tilde{\mu}_{\Theta }$ is the value of the
chemical potential at the tricritical ({\it theta}) point; $\lambda =\lambda
_{0}\Delta \tilde{\mu}$; $a$, $b$, $\lambda _{0}$, and $v$ are
system-dependent parameters. The conditions $h=0$ and $\tilde{\tau}=0$
determine the $\lambda $-line. At the tricritical point, the coefficient $%
\lambda $ changes its sign, being negative along the $\lambda $-line above $%
\Theta $ and positive below $\Theta $.

The equilibrium values $\psi ^{\prime }$ and $\psi ^{\prime \prime }$ of the
order parameter are found from the conditions $(\partial \Delta \tilde{\Omega%
}/\partial \psi )_{T,h}=0$ and $\Delta \tilde{\Omega}(\psi ^{\prime
})=\Delta \tilde{\Omega}(\psi ^{\prime \prime })$. The concentration (volume
fraction) $\phi $ is related to the polymer order parameter $\psi $ by
\begin{equation}
\phi =\left( \frac{\partial \Delta \tilde{\Omega}}{\partial \Delta \tilde{\mu%
}}\right) _{T,h}=a\psi ^{2}-\lambda _{0}\psi ^{4}.  \label{eq_phi_psi2}
\end{equation}
In the limit of infinite degree of polymerization ($h=0$), we find for the
limiting phase-separation boundary shown by the dashed line in Fig. 1:

\begin{equation}
\phi ^{\prime }=0,\qquad \phi ^{\prime \prime }=\frac{\lambda _{0}a}{2v}%
b|\Delta \tilde{T}|\left[ 1-\frac{\lambda _{0}^{2}}{4va}b|\Delta \tilde{T}|%
\right]  \label{phi_zerofield}
\end{equation}
At non-zero $h$, a phase separation (``wing'') critical line emerges,
defined by
\begin{equation}
\psi _{{\rm c}}^{2}(h)=\frac{1}{4}\left( \frac{2h}{v}\right) ^{2/5}\simeq
\phi _{{\rm c}}/a,  \label{psi_c}
\end{equation}
\begin{equation}
T_{{\rm c}}(h)=\Theta \left\{ 1+\frac{5v}{\lambda _{0}ab}\left[ a\psi _{{\rm %
c}}^{2}(h)-3\lambda _{0}\psi _{{\rm c}}^{4}(h)\right] \right\} ^{-1}.
\label{Tc}
\end{equation}
Asymptotically, the ratio of the slopes of the limiting ($h=0$)
phase-separation boundary to the critical ``wing'' line is universal in the
Landau expansion and is equal to $5/2$. A comparison between the results
obtained from the Landau expansion (\ref{exp_phi}) and from the mean-field
Flory model at $N\gg 1$ has shown that the ordering field $h$ can be
identified with the degree of polymerization $N$ as $\left( 2h/v\right)
^{-2/5}\sim \sqrt{N}$. Consequently, the near-tricritical Landau model
satisfies the mean-field scaling given by Eq. (\ref{Widom_scaling}).

The Landau theory does not include fluctuations and does not recover the
3d-Ising limit exhibited by real polymer systems. Therefore, we have
modified the expansion (\ref{exp_phi}) using the crossover procedure based
on the renormalization-group matching method (see \cite{PS:99} and
references therein). The details of the calculations will be published
elsewhere. The key point of the approach is representing the polymer order
parameter as a sum of a regular $\psi _{0}$ and a ``critical'' $\delta \psi $
part $\psi =\psi _{0}+\delta \psi $ and rewriting expansion (\ref{exp_phi})
in terms of $\delta \psi $. The critical part is expressed in terms of the
distance to the critical temperature (at certain field $h$) $\tau =\left[
T-T_{{\rm c}}(h)\right] /T$. The crossover procedure is implemented by
replacing the temperature variable $\tau $ and the order parameter $\delta
\psi $ in the corresponding Landau expansion with renormalized quantities $%
\tau _{{\rm \times }}$ and $\delta \psi _{{\rm \times }}$, respectively,
such that \cite{PS:99}
\begin{equation}
\tau _{{\rm \times }}=\tau Y^{-\alpha /2\Delta _{{\rm s}}},\quad \quad
\delta \psi _{{\rm \times }}=\delta \psi Y^{(2\gamma -3\nu )/4\Delta _{{\rm s%
}}}\,,  \label{ren}
\end{equation}
where $\alpha $, $\gamma $, $\nu $, and $\Delta _{{\rm s}}$ are universal
critical exponents with the following 3d-Ising values adopted in this work: $%
\alpha =0.11$, $\gamma =1.239$, $\nu =0.630$, and $\Delta _{{\rm s}}=0.51$
\cite{GZ:87,PS:99}. The crossover function $Y$ is to be determined from the
equation
\begin{equation}
1-(1-\bar{u})Y=\bar{u}[1+(\Lambda /\kappa )^{2}]^{1/2}Y^{\nu /\Delta _{{\rm s%
}}},  \label{Y}
\end{equation}
where $\bar{u}$, a normalized coupling constant roughly independent of $h$,
and $\Lambda =\Lambda _{0}(2h/v)^{2/5}$, a dimensionless ``cutoff''
wavenumber assumed to be inversely proportional to the radius of gyration $%
R_{{\rm G}}$, are two crossover parameters. The parameter $\kappa $ is
inversely proportional to the correlation length and serves as an effective
distance to the critical point. In the simplest approximation
\begin{equation}
\kappa ^{2}=-2c_{t}\tau Y^{(2\nu -1)/\Delta _{{\rm s}}},  \label{k2_1}
\end{equation}
where the parameter $c_{t}=c_{t0}(2h/v)^{2/5}$ is associated with the
amplitude $\bar{\xi}_{0}$ of the mean-field correlation length $\bar{\xi}$.
Close to the critical point $Y\rightarrow (\kappa /\bar{u}\Lambda )^{\Delta
_{{\rm s}}/\nu }\rightarrow 0$ and the thermodynamic properties exhibit
3d-Ising asymptotic behavior. Far away from the critical point, $%
Y\rightarrow 1$, and the mean-field expansion (\ref{exp_phi}) is recovered.
The crossover temperature (``Ginzburg number'') $\tau _{0}\sim (\bar{u}%
\Lambda )^{2}/c_{t}=[(\bar{u}\Lambda _{0})^{2}/c_{t0}](2h/v)^{2/5}\sim \phi
_{{\rm c}}$ vanishes at the {\it theta} (tricritical) point. The physical
origin of the crossover to tricriticality is a competition between the
radius of gyration $R_{{\rm G}}$ and the correlation length $\xi $, since $%
\Lambda /\kappa \sim \xi /R_{{\rm G}}$, while the parameter $\Lambda
^{2}/c_{t}\sim (\bar{\xi}_{0}/R_{{\rm G}})^{2}$ defines the crossover
temperature $\tau _{0}$. Specific $N$-dependencies of the Ising critical
amplitudes, predicted by de Gennes' scaling \cite{Sa:89,deGennes}, can be
also obtained from our theory with the assumption $(2h/v)^{-2/5}\sim \sqrt{N}
$.

We have applied the renormalized (crossover) Landau model to describe the
experimental data \cite{DK:80,XS:96,NC:78} on phase separation in polymer
systems and have obtained excellent agreement (solid lines in Figs. 1-3).
The description of all the systems with a variety of degrees of
polymerization requires only four non-universal parameters, namely $a$, $%
\lambda _{0}$, $C=\lambda _{0}a/v$, and the ``bare'' crossover temperature $%
\sim (\bar{u}\Lambda _{0})^{2}/c_{t0}$, which do not depend on\ molecular
weight. Moreover, the combination $(\bar{u}\Lambda _{0})^{2}/c_{t0}$ and $%
AC^{-\beta }$ can be taken to be the same within the available experimental
resolution not only for different molecular-weight samples, but also for
different substances. This feature makes the solid curve in Figs. 3 and 4
truly universal for all systems studied.

The universality demonstrated in Fig. 4 requires both $\tau $ and $\phi _{%
{\rm c}}$ to be small. In first approximation some non-asymptotic effects
are incorporated into the universal scaling description. The slight
dependence of $A$ on $\phi _{{\rm c}}$ for moderate molecular weights (at
larger $\phi _{{\rm c}}$), shown in Fig. 1, absorbs non-asymptotic
corrections to the critical limit. A non-asymptotic (at larger $\tau $)
non-linearity of the phase separation boundary in the tricritical (zero
field) limit can be accounted for by a term quadratic in $\tau $.

Renormalization-group calculations \cite{S:75} have shown the existence of
logarithmic corrections to mean-field tricriticality: the coefficients $v$
and $\lambda _{0}$ in expansion (\ref{exp_phi}) are renormalized, so that
the critical line has zero slope at the {\it theta}-point. The resolution of
the existing experimental data is not sufficient to convincingly determine
the logarithmic corrections: the description is equally good with or without
the corrections.

We acknowledge valuable discussions with M. E. Fisher, S. C. Greer, A. Z.
Panagiotopoulos, J. V. Sengers, and B. Widom. The research at the University
of Maryland was supported by DOE Grant No. DE-FG02-95ER-14509.

\begin{center}
\newpage

{\large Figure Captions}
\end{center}

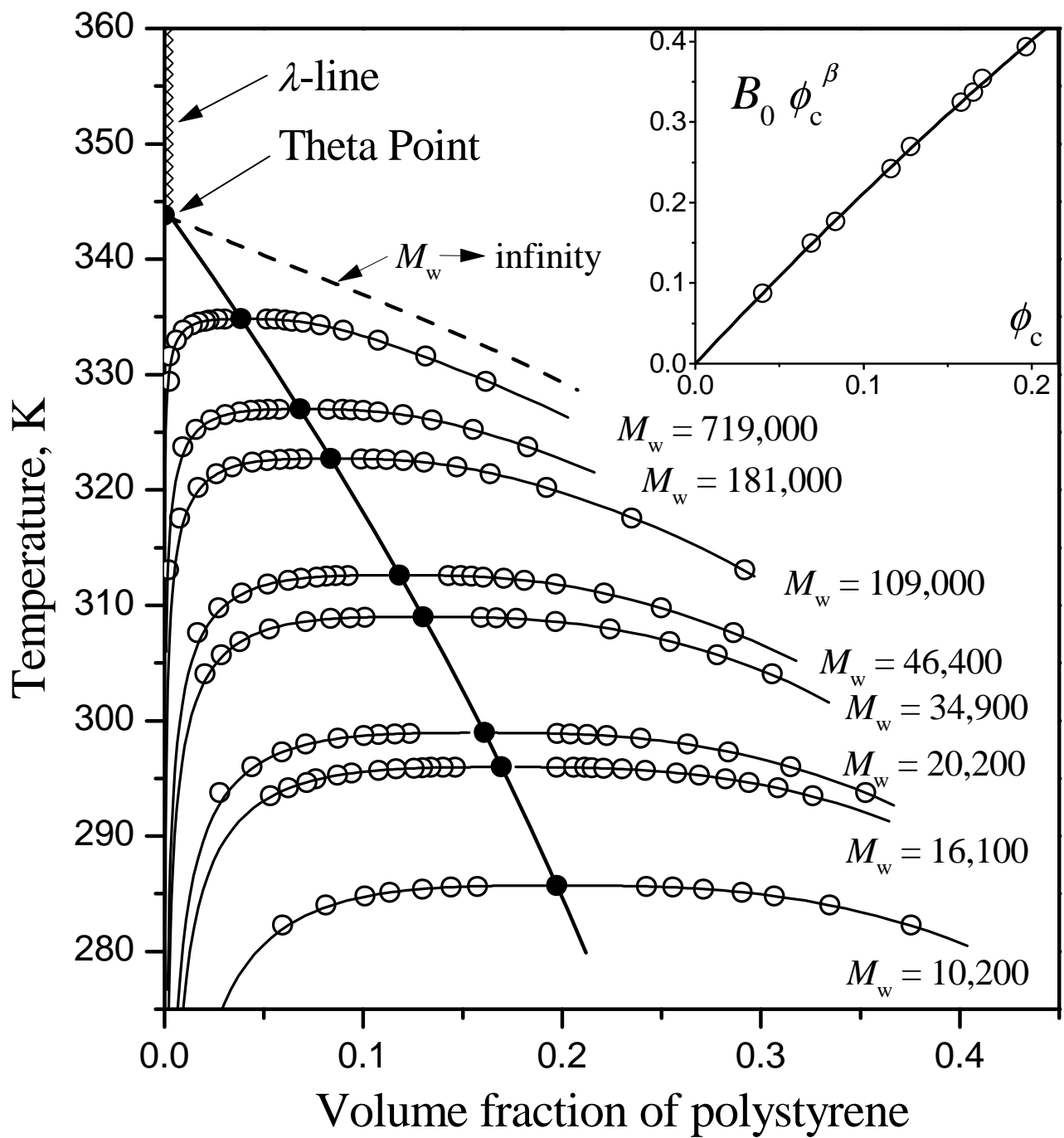
Figure 1. Phase-coexistence curves for solutions of polystyrene of various
molecular weights $M_{{\rm w}}$ in methylcyclohexane. Symbols indicate
experimental data by Dobashi {\it et al.} \cite{DK:80}. The insert shows
dependence of the critical amplitude $B_{0}$ on the critical concentration.
Solid curves represent the crossover theory.

\bigskip

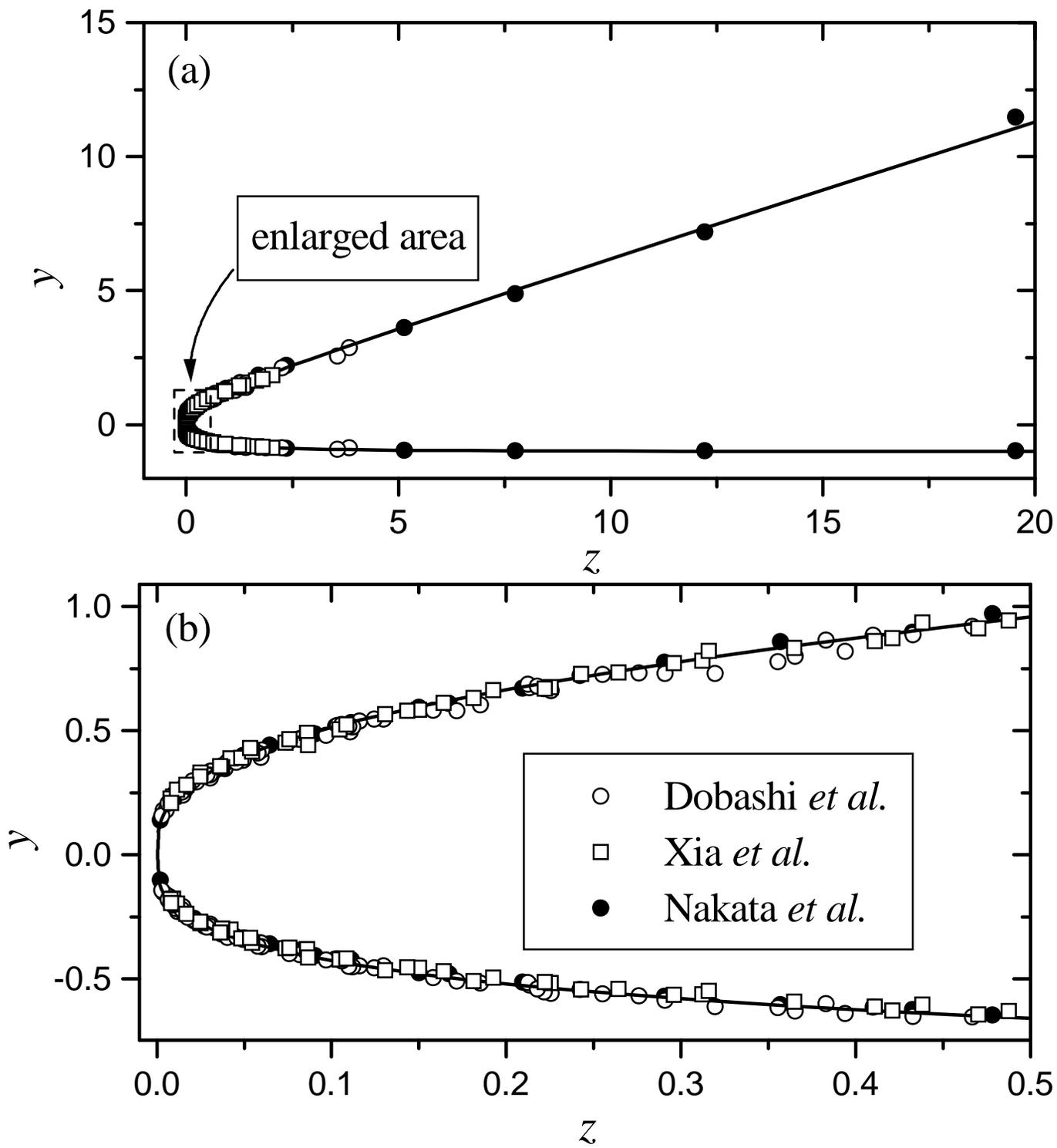
Figure 2. Universal scaled coexistence curve of polymer solutions: (a) the
entire range, (b) the critical region. Solid line is calculated from the
crossover theory.

\bigskip

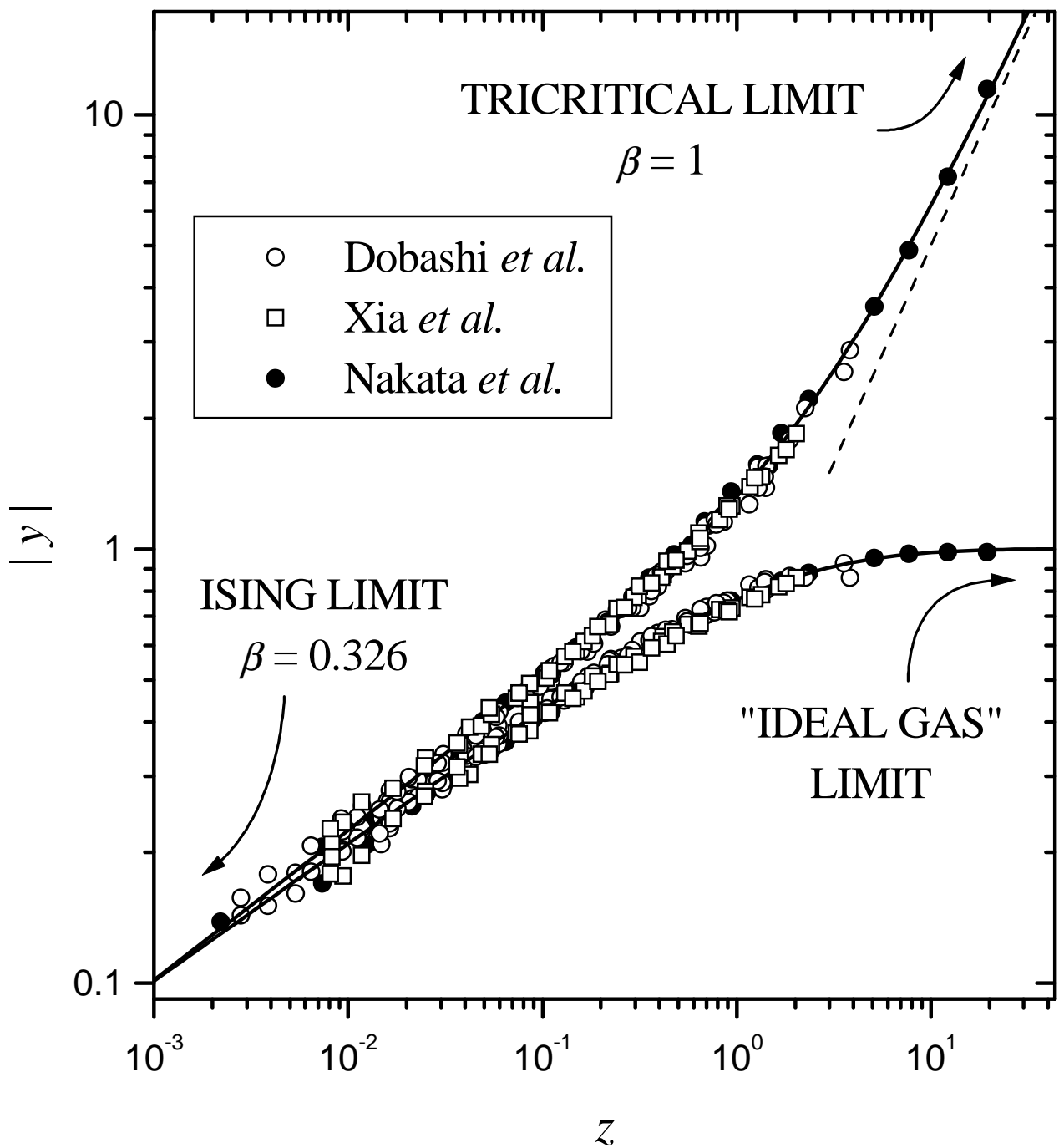
Figure 3. Universal scaled coexistence curve of polymer solution in a
double-logarithmic scale showing crossover from Ising behavior to ``ideal
gas'' and tricritical ({\it theta}-point) behavior. Solid line is calculated
from the crossover theory.

\end{document}